\documentclass[aps,prb,twocolumn,groupedaddress,showpacs,10pt]{revtex4-1}

\usepackage{amssymb,amsfonts,amsmath,latexsym,epsfig,subfigure,bm}
\usepackage[colorlinks=true,citecolor=blue,urlcolor=blue]{hyperref}

\newcommand{\beq}{\begin{equation}}
\newcommand{\eeq}{\end{equation}}

\renewcommand{\bf}{\mathbf}
\def\d{\partial}

\def\tr{\mathrm{tr}}

\def\cH{\mathcal{H}}
\def\r{\bf{r}}
\def\k{\bf{k}}
\def\p{\bf{p}}
\def\s{\bf{s}}

\def\rf#1{(\ref{#1})}

\begin{document}

\title{Antiferromagnetic topological insulators in cold atomic gases}
\author{Andrew M.~Essin}
%\author{Ana Maria Rey}
\author{Victor Gurarie}
\affiliation{Department of Physics, CB390, University of Colorado,
Boulder CO 80309, USA}
%\email[]{Your e-mail address}
%\homepage[]{Your web page}
%\thanks{}
%\altaffiliation{}
%\noaffiliation
\date{\today}

\begin{abstract} 
We propose a spin-dependent optical lattice potential that realizes
a three-dimensional  antiferromagnetic topological insulator in a gas of cold, two-state
fermions such as alkaline earths, as well as a model that 
describes the tight-binding limit of this potential.  We discuss the
physically observable responses of the gas that can verify the presence
of this phase, in particular rapid rotation in response to the trap potential. We also point out how this model can be used to obtain two-dimensional flat bands with nonzero Chern number. 
\end{abstract}

\maketitle

\section{Introduction}

The use of cold atomic gases to implement many-body models of condensed matter physics is by now well-advanced. 
%A number of such models have already been implemented including models of weakly and strongly interacting Bose and Fermi gases, a model of bosons on a lattice undergoing superfluid to Mott insulator transition, 
%and a Hubbard model of fermions on a lattice with SU(2) (and sometimes even SU(N)) symmetric spin. 
The goal of this research is twofold: to simulate existing materials with cold atoms and to manufacture 
Hamiltonians 
%which so far have not been 
unseen in solids.
%the condensed matter context. 
 
A particularly strong effort in the field over the last decade has been directed towards recreating the integer and 
%perhaps 
fractional quantum Hall effects with cold atoms by simulating an orbital magnetic field for neutral atoms, achieving slow but steady progress. The quantum Hall effects realize a large variety of topological states of matter. Not all of them have been unambiguously seen in semiconductor heterostructures, and some of those not yet
obtained may be important for applications~\cite{Nayak2008}.
One hopes that quantum Hall effects with cold atoms will provide %us with 
ways to investigate those states experimentally.
%A slow but steady progress has been achieved along this line of research.
% Recently, for example, lattices with fluxes of alternating signs were created by imprinting phases onto the tunneling matrix elements of atoms hopping on a square lattice, as a first step towards realizing 
%lattices with magnetic  fluxes of the same sign, ultimately  leading to integer quantum Hall effect. 

Parallel to that effort, a number of breakthroughs in condensed matter physics in recent years have led to an understanding that the integer quantum Hall effect is but one particular system in a class of noninteracting fermionic systems, in a variety of spatial dimensions, which have received the name of topological insulators (TI)~\cite{HasanKane2010}. All TIs have gapped bulk and gapless edge states, and respond to external electromagnetic perturbations in a quantized way~\cite{Qi2008,Ryu2010}.  
The quantum Hall effect is confined to two dimensions, but three-dimensional (3D) topological insulators have also been observed experimentally. In each spatial dimensionality there are five distinct types of topological insulators and superconductors~\cite{Ryu2010}, with only two of those five seen experimentally in 3D in real materials.  Cold atoms 
may end up providing the only way to manufacture the 3D TIs not yet seen.  

It is also suspected that in the presence of interactions TIs may develop phases similar in some sense to those of the fractional quantum Hall effect~\cite{Levin2009,*Karch2010,*Swingle2011}.  Theoretical study of these interacting phases is currently a rapidly developing subject. While it is not yet known if these phases can be seen in a condensed matter context, it is natural to consider TIs with cold atoms, whose interactions can often be controlled or chosen in advance. 

%It is especially interesting to consider the implementation of the 3D strong (TI).~\cite{FuKaneMele2007,MooreBalents2007,Roy2009}  
The distinctive signature of  3D  TIs, in addition to gapless, Dirac-type excitations localized at the surfaces of
the system, is a strong, quantized magnetoelectric response.  The former
is best seen as the 3D counterpart to the chiral edge 
states of the integer quantum Hall phase, and the latter as the 
counterpart of the %quantized 
Hall conductance of that phase.

We propose a model which realizes an unusual (and thus far unseen) 3D TI, called the antiferromagnetic topological insulator (AFTI). 
The AFTI bulk is similar to
%This insulator belongs to the same type as 
the standard strong TI described by Refs.~\onlinecite{FuKaneMele2007,MooreBalents2007,Roy2009}.  However, time-reversal invariance, crucial to
%a crucial symmetry of 
that type of a TI, is implemented in a different way,
with the result that the magnetoelectric response of the TI becomes a 
ground-state property in
% leading to a distinct manifestation of a magneto-electric response more suitable for observation with cold atoms as the role of the applied external magnetic field is played by any external {\sl scalar} potential, including
the trap potential always present in cold atomic setups.  As a result, in response to the applied trap potential this system starts rotating rapidly, a tell-tale signature that we hope can be used to detect this phase.  

The model we propose can be implemented in cold atoms by an extension of the idea of artificial gauge fields~\cite{Spielman2009}. The construction 
%we propose 
involves atoms with only two internal states, and we hope not only that the model proposed here possesses features (magnetoelectric response to an applied scalar potential) making it more suitable for observation and study with cold atoms, but also that this provides a simplification compared to existing schemes to implement strong TIs~\cite{Spielman2010,BeriCooper2011}, as we elaborate below.

Moreover, a tight-binding limit of this model acquires sublattice symmetry and is a chiral 3D TI~\cite{Hosur2010}.  This is a type of 3D insulator distinct from the standard strong TI and, like the AFTI, not yet seen experimentally in solids.

Finally, we propose to use this insulator as a way to create two-dimensional flat bands (surface bands of this insulator) with nonzero Chern number, which are known to have the potential to enhance interaction effects and therefore
aid the formation of fractional quantum Hall states without strong magnetic fields. 

\section{Antiferromagnetic TI}

%The crucial symmetry of a TI is time reversal, 
%in particular 
A starting point towards constructing a strong TI is an identification of a time reversal operation $\mathcal{T}$ that satisfies 
$\mathcal{T}^2 = -1$.  If it is a symmetry, it ensures that the 
energy eigenstates come in degenerate Kramers pairs, which provides the 
simplest way to understand that the edge spectrum must be gapless.

Realization of the TI requires a minimum of four
distinct states per wave vector $\k$.  First, there needs to be some 
degree of freedom on which the symmetry $\mathcal{T}$ can be 
realized, which requires two states that we call spin.  
%We can imagine for 
%definiteness that these are the ground $^3$S$_0$ and excited $^3$P$_0$ states of fermionic
%alkaline-earth-like atoms such as Sr or Yb, which is attractive due to the 
%extremely long lifetime of the excited state and the fact that these can be coupled directly by a laser operating at an optical frequency.  
With just these
two states, however, there will not be a gap in the bulk band structure,
so to achieve a band insulator there must be more states.  The minimal
implementations considered so far require four spin states, as in the
proposal of Ref.~\onlinecite{BeriCooper2011}.

In an effort to minimize the number of internal states
that need to be manipulated, we propose instead to use two sublattices, limiting the number of internal (spin) states to two.  
Furthermore, we want to find a model on the simplest lattice possible,
so we %limit ourselves 
restrict to nearest-neighbor hopping 
terms; this rules out the diamond-lattice tight-binding model of Ref.~\onlinecite{FuKaneMele2007}, 
%on the diamond lattice, 
for example.

%\subsection*{AFTI}

The simplest approach to achieve the physics of the 
TI % that we were able to find  
that we have found realizes the AFTI~\cite{Mong2010}.  The prototype of 
such a system involves electrons that have a Zeeman coupling to an 
N\'{e}el order parameter.  This obviously breaks $\mathcal{T}$
%time-reversal symmetry 
because the order parameter flips under the action of time reversal,
but the magnetic order may be such that the symmetry is restored after an 
appropriate translation.  We represent this translation by a unitary 
operator $T_{1/2}$, for translation through half the magnetic unit cell; 
%if we represent the action of $\mathcal{T}$ by an antiunitary operator 
%$\Theta$, 
the symmetry is implemented by the antiunitary operator
%\beq
$S = \mathcal{T} T_{1/2}$,
%\eeq
which satisfies $S^2 = -1$ and therefore has the crucial property 
necessary for nontrivial topological physics.

An AFTI has two distinct types of surface, called ``antiferromagnetic" and 
``ferromagnetic" in this context.  The first type preserves the symmetry $S$ of the bulk, and therefore supports gapless surface states of Dirac type 
just like a surface of the usual TI.  Ferromagnetic 
surfaces break the symmetry, which opens a gap, and the surface 
realizes a half-quantum-Hall effect, just like a surface of the usual TI
with an added $\mathcal{T}$-breaking perturbation~\cite{Qi2008,Essin2009}.

We have found a spin-dependent, optical lattice potential that realizes
the AFTI, as well as a tight-binding model for the deep-well limit of 
this potential. 
% It may be possible to realize the tight-binding model
%directly by imprinting matrices onto the tight binding matrix elements or with  ``nonabelian gauge fields.''\cite{Jaksch2003} \textbf{others?}  
The tight-binding model has
an extra chiral symmetry which is interesting in its own right~\cite{Ryu2010}, 
%but
%we limit the discussion to effects that can be seen even without that
%symmetry, 
and we display band structures in 
Fig.~\ref{fig:bandstructures} both in the chiral, tight-binding limit and far from it so generic features are distinguishable.

%\section*{Hamiltonians}

\subsection{Optical lattice}

%We propose 
The following spin-dependent, noninteracting Hamiltonian %to 
realizes an AFTI:
{ \allowdisplaybreaks
\begin{align} \label{eq:topins}
H_{AF}(\p,\r,\s) &= \frac{p^2}{2m} + V(\r) + \bf{B}_Z(\r) \cdot {\bm \sigma} ,
\notag\\
V(\r) & = V \left[ \cos q \hat{\bf{x}} \cdot \r + \cos q \hat{\bf{y}} \cdot \r + \cos q \hat{\bf{z}} \cdot \r \right] \notag\\
\bf{B}_Z(\r) &= B_Z 
\sum_{i=1}^4 \bf{b}^i \cos \left(  q\bf{b}^i \cdot \r \right) ,
\end{align}
}
where $q=2\pi/a$ sets the length scale.  Here $\p$ and $\r$ are the 
single-particle momentum and position; $\hat{\bf{x}}$, $\hat{\bf{y}}$, and $\hat{\bf{z}}$ are orthogonal unit vectors; 
% with $\hat{\bf{x}}$, $\hat{\bf{y}}$, and $\hat{\bf{z}}$ the 
% corresponding unit vectors; 
and  $\bm \sigma$ represents the vector of Pauli matrices. Finally, the tetrahedral vectors ${\bf b}^i$ are defined as
$\bf{b}^1 = (-\hat{\bf{x}}+\hat{\bf{y}}+\hat{\bf{z}})/2$, 
$\bf{b}^2 = (\hat{\bf{x}}-\hat{\bf{y}}+\hat{\bf{z}})/2$, 
$\bf{b}^3 = (\hat{\bf{x}}+\hat{\bf{y}}-\hat{\bf{z}})/2$, 
$\bf{b}^4 = -(\hat{\bf{x}}+\hat{\bf{y}}+\hat{\bf{z}})/2$.

\begin{figure} 
\subfigure[]{ \includegraphics[width=0.3\columnwidth]{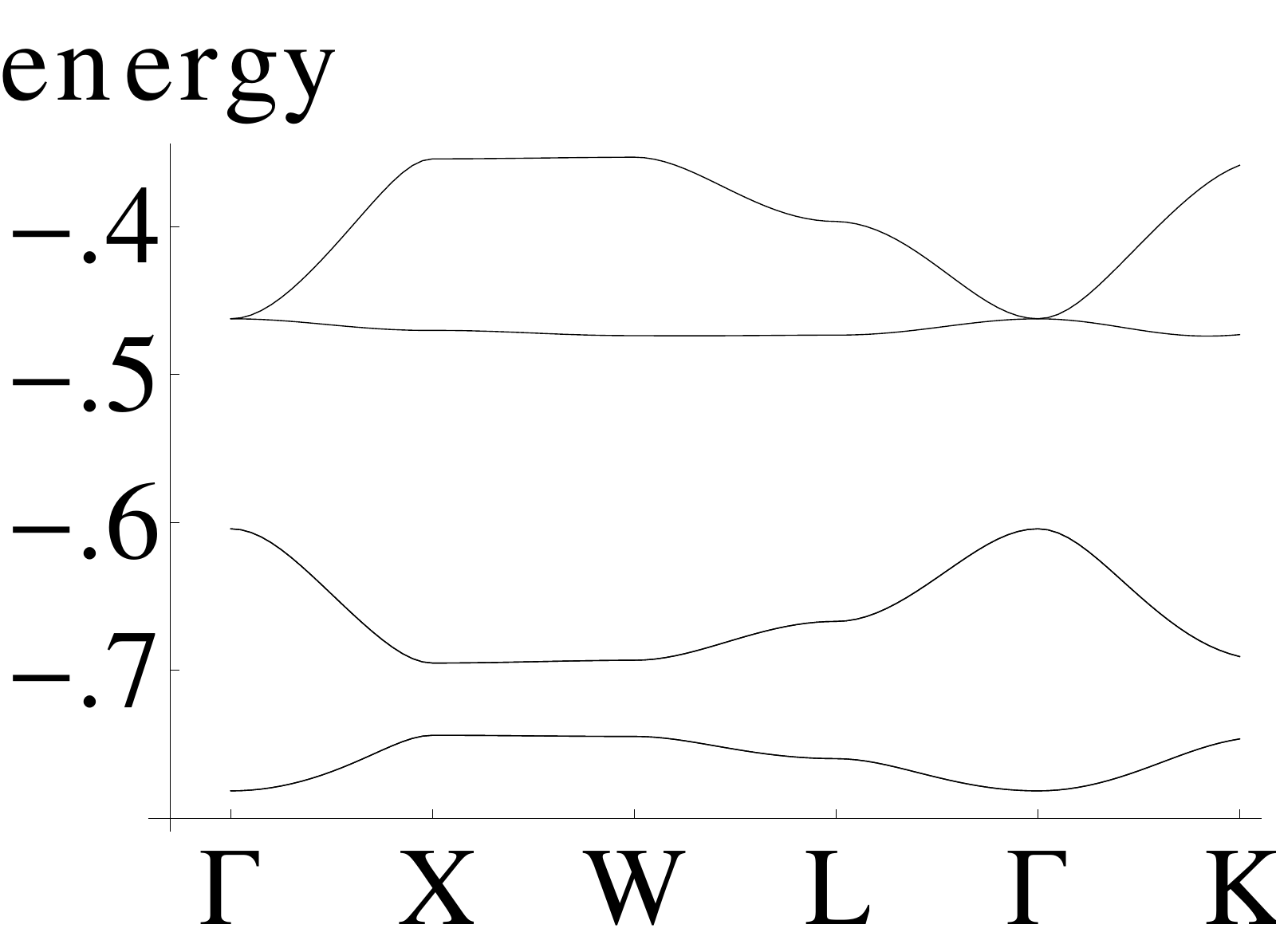}
\label{fig:bulkbands}
} \hspace*{-0.1in}
\subfigure[]{ \includegraphics[width=0.3\columnwidth]{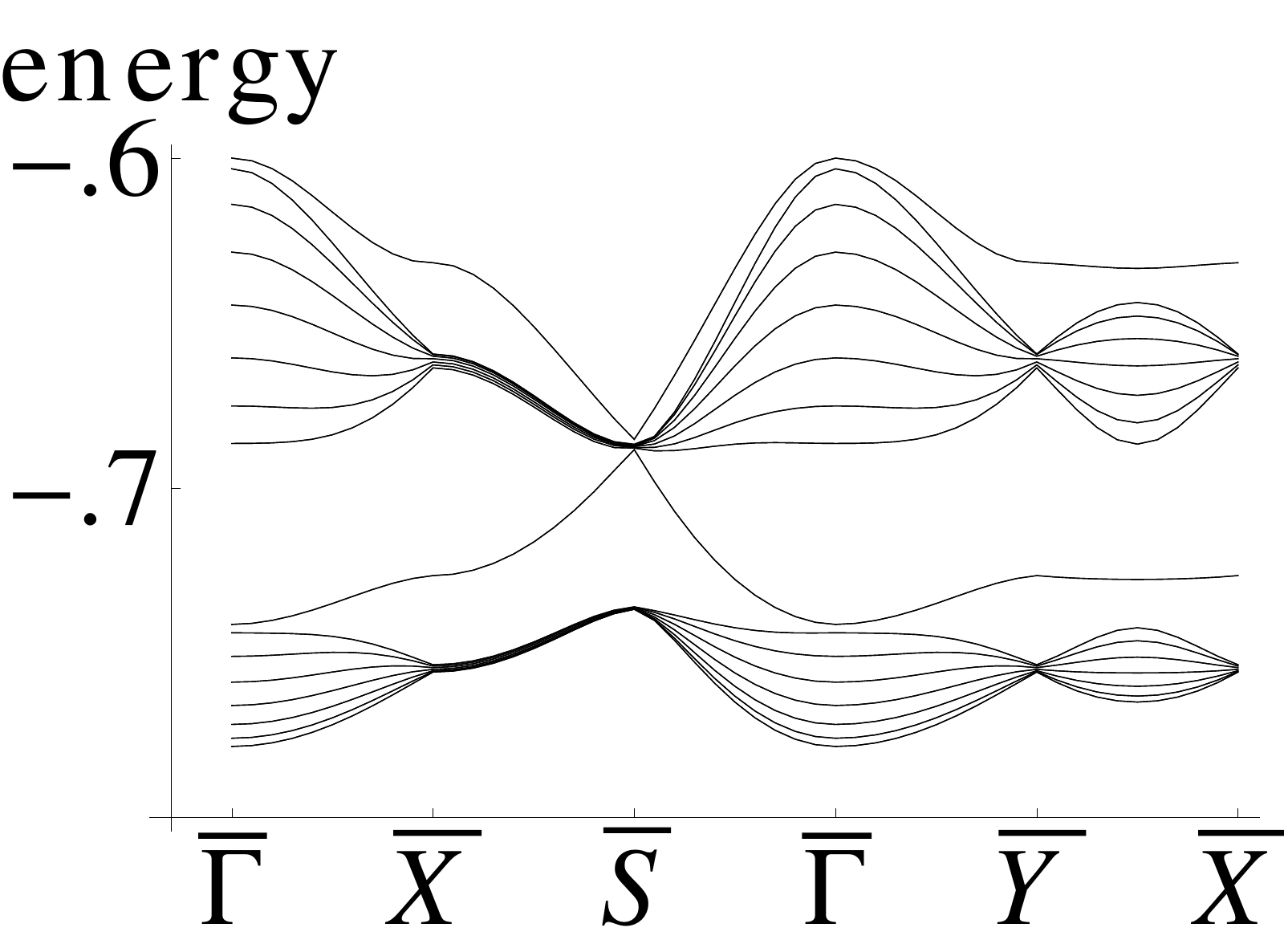}
\label{fig:ss}
} \hspace*{-0.1in}
\subfigure[]{ \includegraphics[width=0.3\columnwidth]{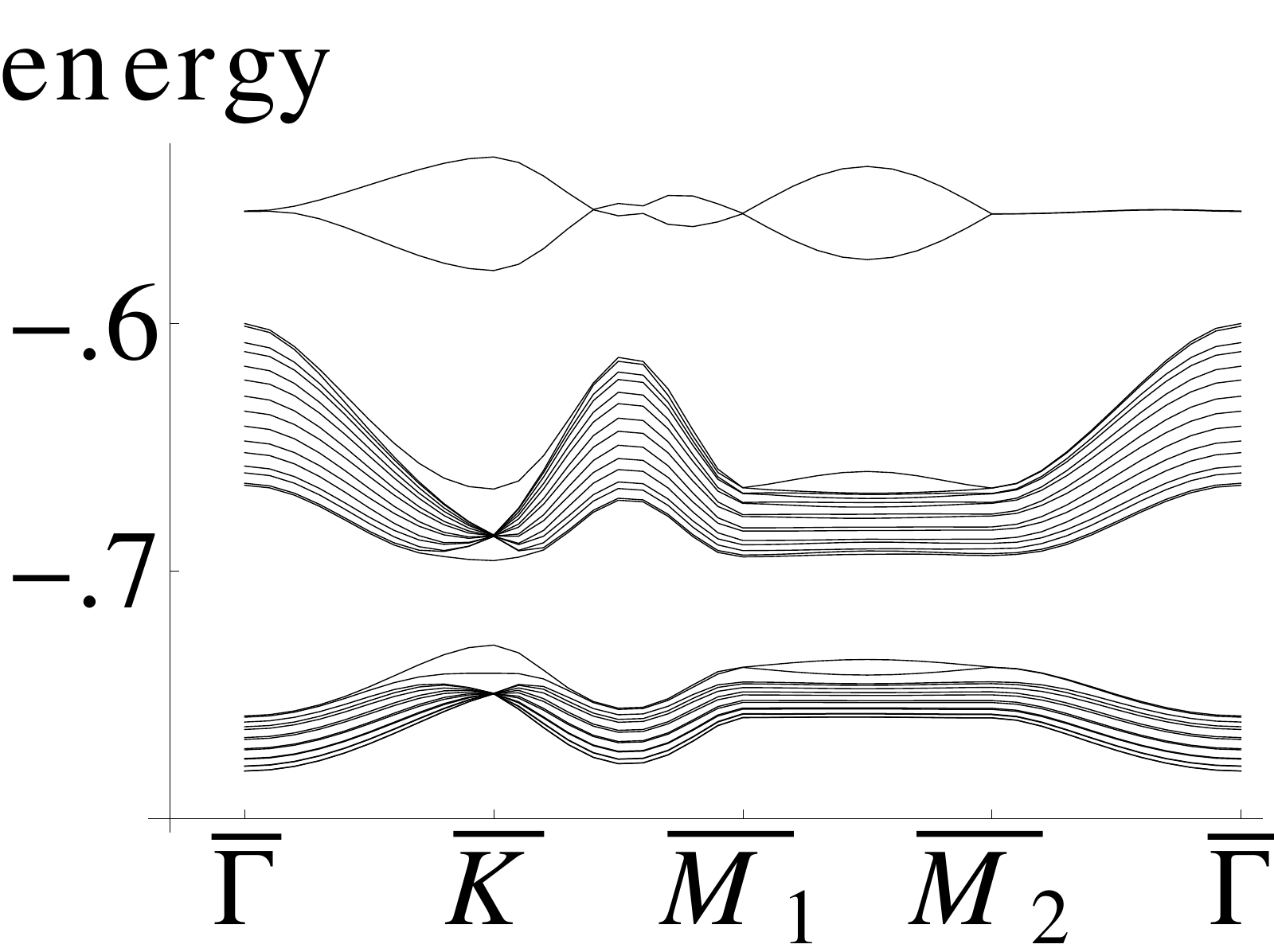}
\label{fig:hs}
} 
\\
\vspace*{-0.15in}
\subfigure[]{ \includegraphics[width=0.3\columnwidth]{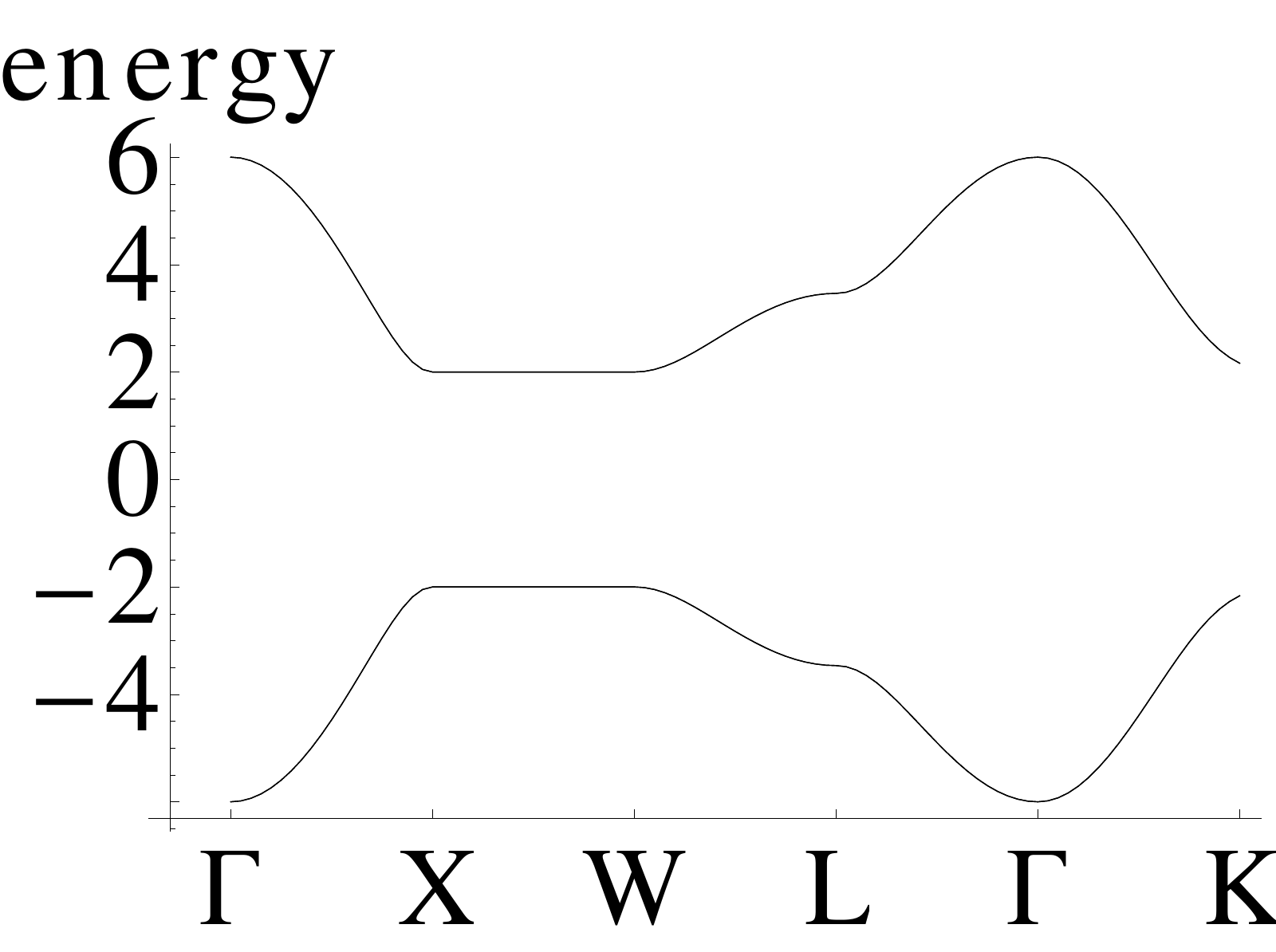}
\label{fig:tbbulkbands}
} \hspace*{-0.1in}
\subfigure[]{ \includegraphics[width=0.3\columnwidth]{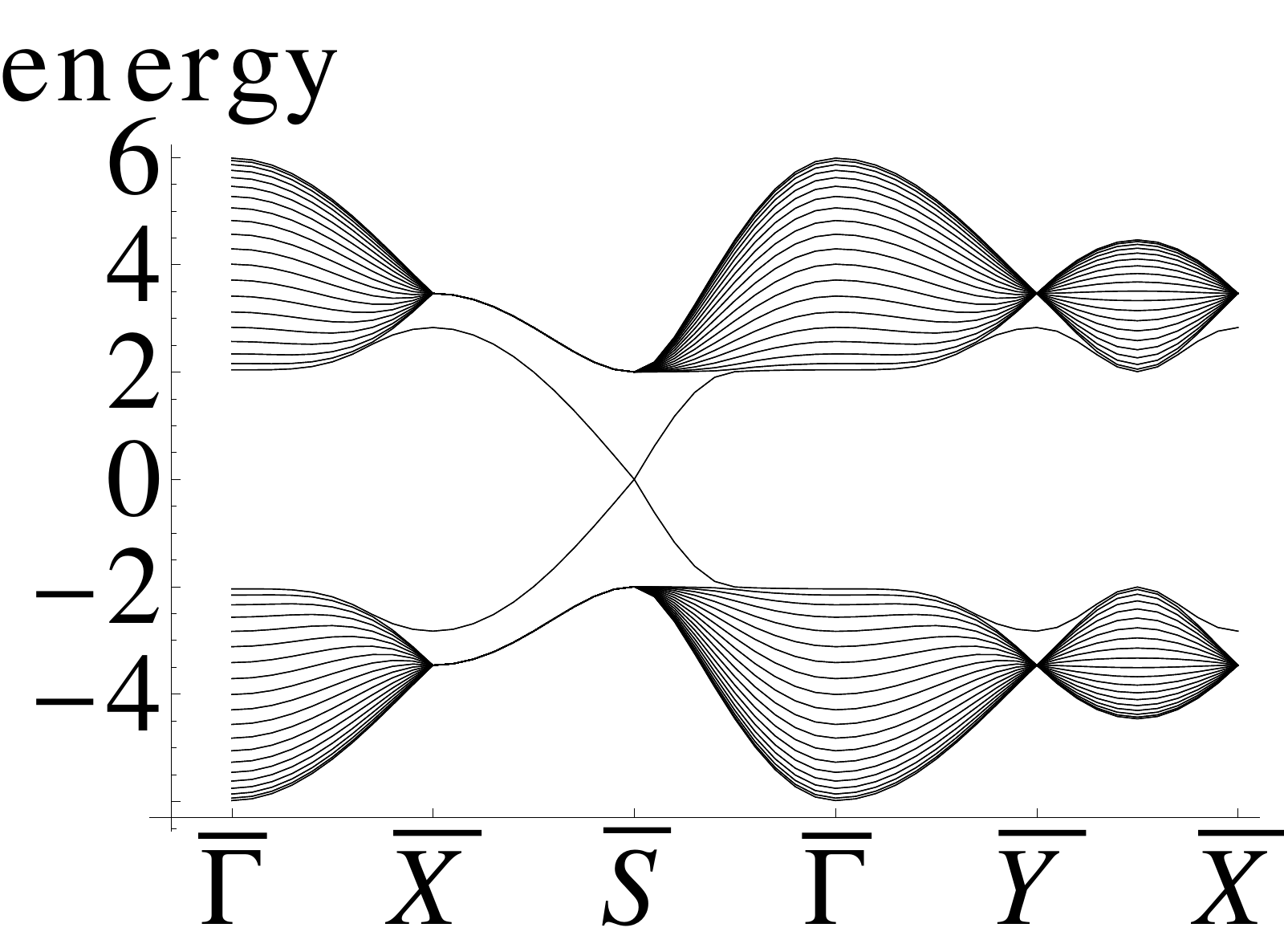}
\label{fig:tbss}
} \hspace*{-0.1in}
\subfigure[]{ \includegraphics[width=0.3\columnwidth]{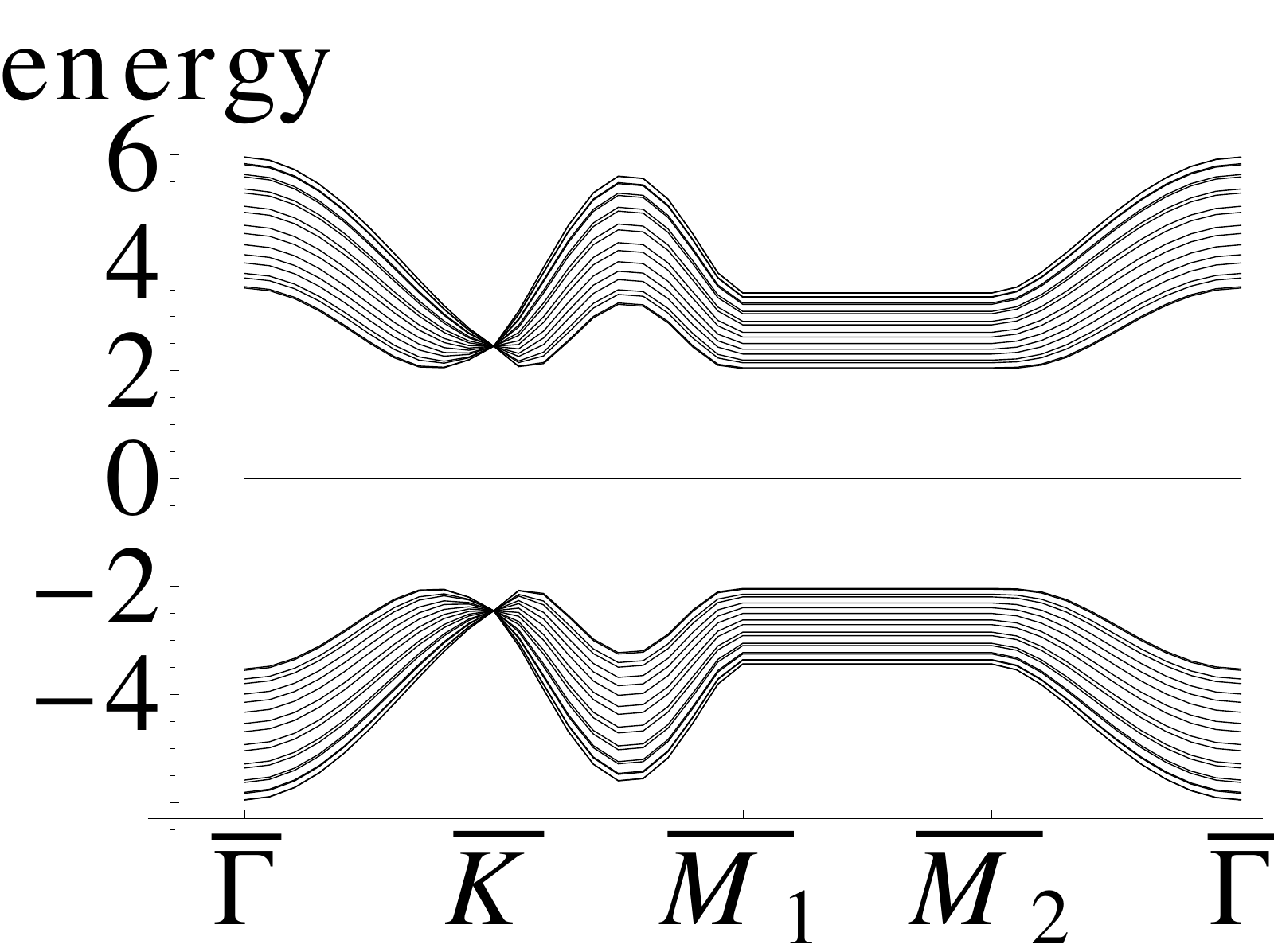}
\label{fig:tbhs}
}
\\
\vspace*{-0.1in}
\subfigure[]{ \includegraphics[width=0.15\columnwidth]{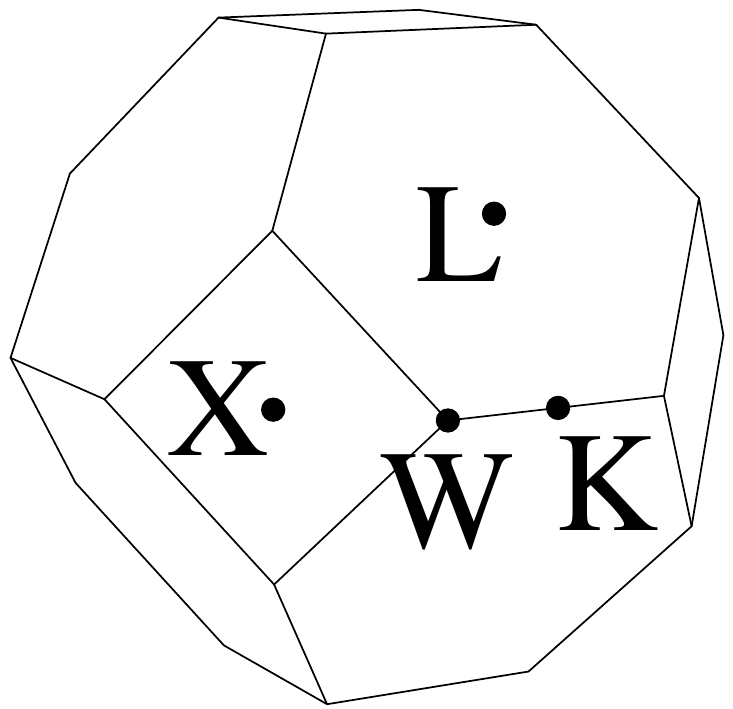}
\label{fig:bz} 
} 
\subfigure[]{ \includegraphics[width=0.15\columnwidth]{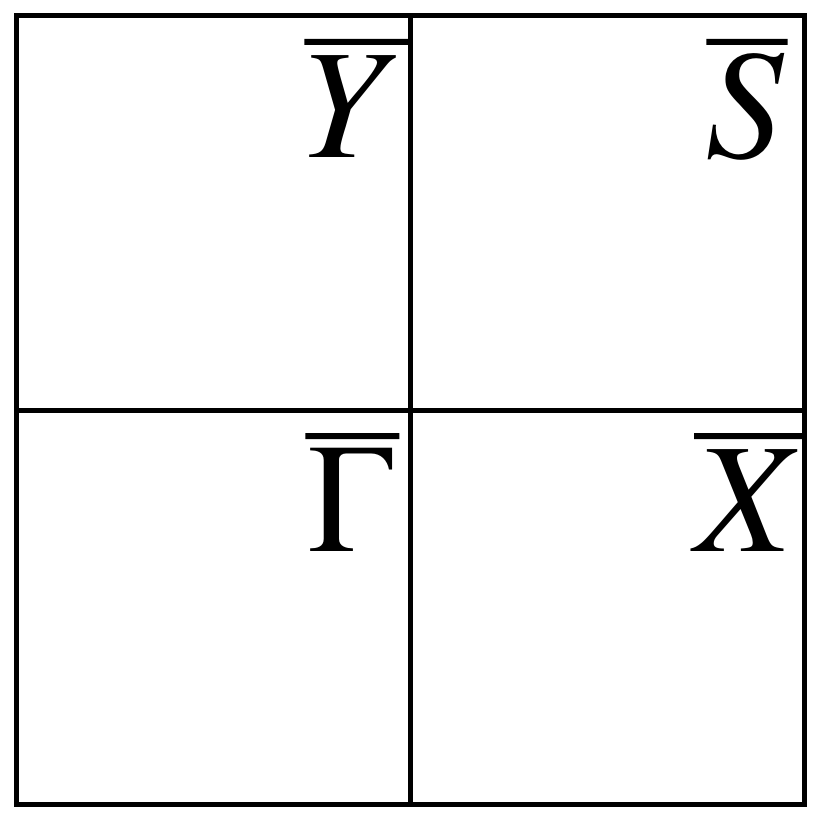}
\label{fig:sbz}
} 
\subfigure[]{ \includegraphics[width=0.15\columnwidth]{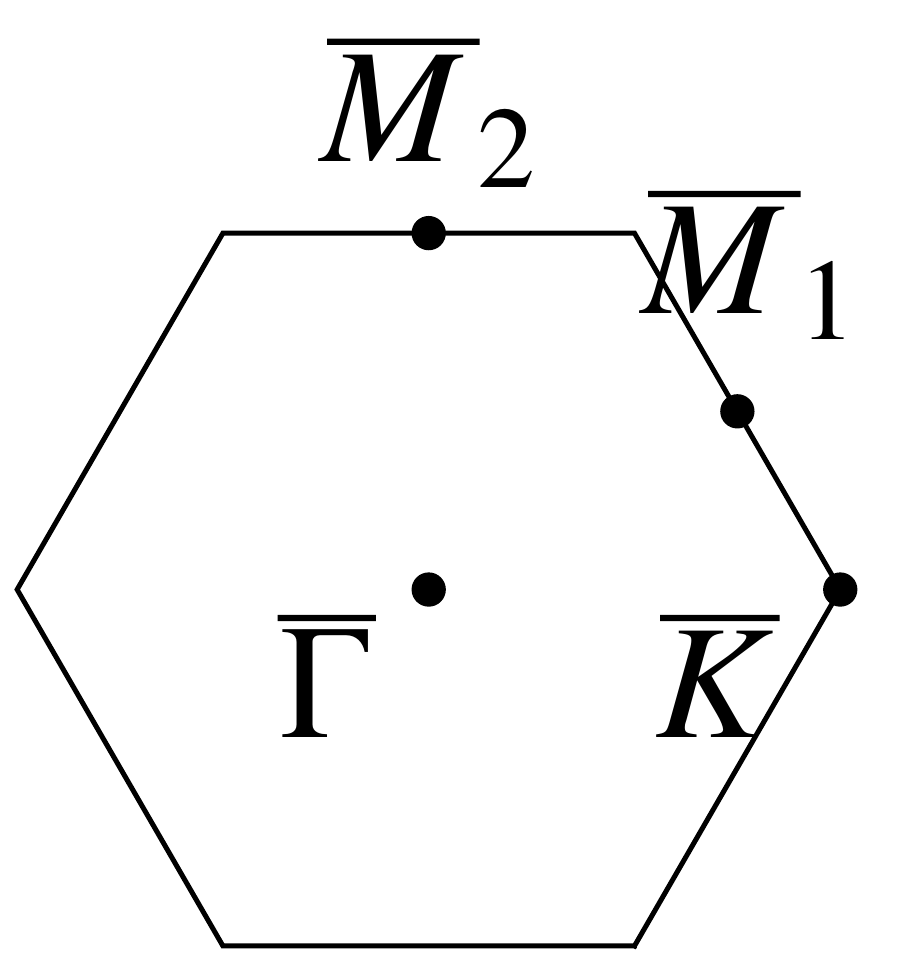} 
\label{fig:hbz}
} 
\subfigure[]{ \includegraphics[width=0.3\columnwidth]{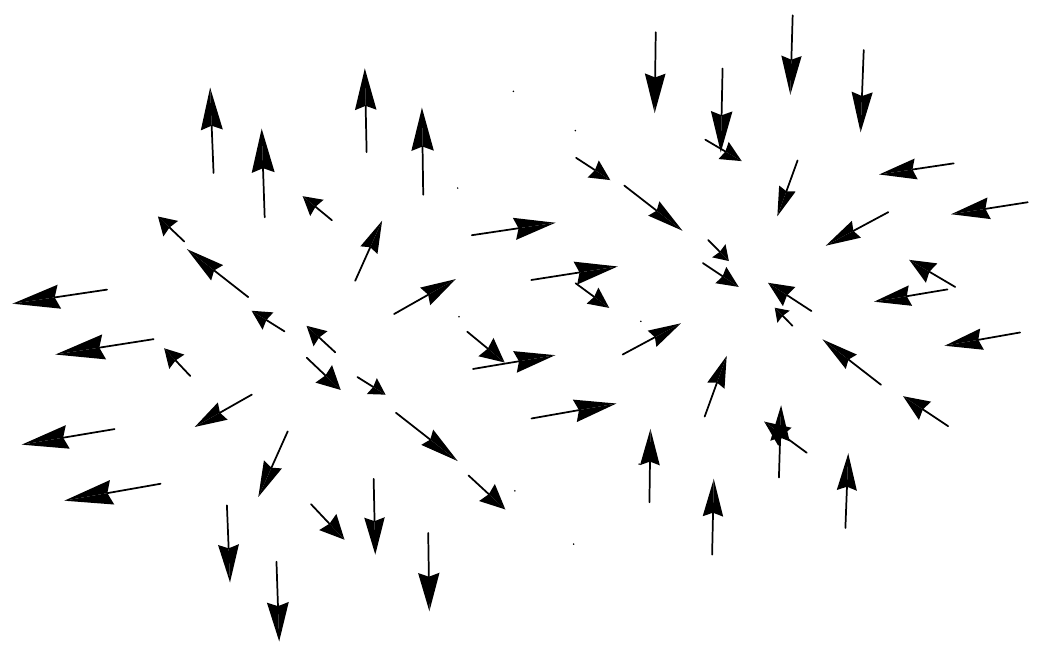}
\label{fig:texture}
}
\caption{ \label{fig:bandstructures} 
The spectra of Eq.~\rf{eq:topins} in (a) bulk, (b) a (100) slab, and 
(c) a (111) slab, with parameters $B_Z = 3 V/2 = \hslash^2 q^2/2m = 1$; 
and corresponding geometries for Eq.~\rf{eq:tb} [(d), (e), (f)] with 
parameters $t = t_M = 1$. 
(g) The bulk Brillouin zone; the zone center $\Gamma$ is not shown.
(h,i) The (100) and (111) surface Brillouin zones.
(j) $\bf{B}_Z(\r)$.
}
\end{figure}

The potential 
$V$ creates a spin-independent, cubic lattice, while the Zeeman field 
$\bf{B}_Z(\r)$,  a sum of four one-dimensional, spin-dependent terms, creates an alternating magnetic ``hedgehog'' texture around 
the wells of that lattice
[see Figure \ref{fig:texture}].  
This is the NaCl structure, which has the 
translation symmetry of a face-centered-cubic (fcc) lattice.

The Zeeman field $\bf{B}_Z(\r)$ breaks %time-reversal symmetry 
$\mathcal{T}$ since $\bm{\sigma}=-\sigma_y \bm{\sigma}^* \sigma_y$, but the symmetry is restored by a translation $T_{1/2}$ through $a$ 
(along any of the cubic axes).
%, thereby interchanging the 
%sublattices.  
This Hamiltonian therefore has the symmetry $S$ described earlier,
%in the introduction, 
which enables a topologically nontrivial phase.  

This Hamiltonian is gapped at a filling of one particle for every well
of $V(\r)$, which is two particles per unit cell of $\bf{B}_Z$; %as seen in 
see Figure \ref{fig:bulkbands}.  (Each band is doubly degenerate since 
the combination of $S$ and inversion is a symmetry; see below
for inversion symmetry). In other words, $\bf{B}_Z$ 
%creates an insulating state out of 
gaps the simple cubic metal described by $p^2/2m + V$.
The resulting insulator is topologically nontrivial, which is computed 
most simply as follows.

In addition to the symmetry $S$, this Hamiltonian also possesses inversion
symmetry $(\p,\r) \rightarrow -(\p,\r)$.  This allows us to 
compute the bulk topological invariant, Eq.~(6) of
Ref.~\onlinecite{Turner2010}, in terms of inversion parities at the eight
inversion-symmetric wave vectors $\bm{\Gamma} = 0$, 
$(\bf{X}^1,\bf{X}^2,\bf{X}^3) = \pi(\hat{\bf{x}},\hat{\bf{y}},\hat{\bf{z}})$, and 
$\bf{L}^i = \pi\bf{b}^i$.  We find that at $\bm{\Gamma}$ both filled
bands are inversion-even, at $\bf{X}^i$ both are inversion-odd, and at 
$\bf{L}^i$ there are one of each parity.  The total number of inversion-odd
states is 10, which is twice an odd number; therefore the band
structure is topologically nontrivial.  (Note that the prescription of 
Ref.~\onlinecite{FuKane2007,*Hughes2011} does
not apply to this system since the Kramers pairs are not degenerate with 
respect to inversion at $\bf{L}^i$).

In this potential, a surface normal to $\hat{\bf{x}}$, $\hat{\bf{y}}$, or
$\hat{\bf{z}}$ [a (100)-type surface] retains the symmetry $S$ of the 
bulk, and so is an antiferromagnetic surface in the terminology of 
Ref.~\onlinecite{Mong2010}.  Such a surface possesses gapless edge states,
as seen in Figure \ref{fig:ss}.  By contrast, a surface normal to 
$\bf{b}^i$ [a (111)-type surface] breaks the symmetry and is gapped, as
in Figure \ref{fig:hs}.  These surfaces support a half-quantum-Hall 
effect.

\subsection{Tight-binding model}

In the deep-well limit, the Hamiltonian \rf{eq:topins} reduces to
% It is perhaps easiest to think about the physics of the above potential in 
%terms of
the following tight-binding model on the cubic lattice %, 
%or really the rock salt (NaCl) lattice, 
with nearest-neighbor, spin-dependent 
%and 
%-independent 
hopping terms (both $t$ and $t_M$ are real):
%[see Figure \ref{fig:tbtexture}]:
\begin{align} \label{eq:tb}
\hat{H}_{tb} &= \!\sum_{\r \in A}\! \sum_{\bf{e}} \hat{c}_{\r}^\dag 
\left[ t +  t_M \bf{e} \cdot \bm{\sigma} \right]  \hat{c}_{\r+\bf{e}}
+ \mathrm{H.c.}
= \sum_{\r,\r'} \hat{c}_{\r}^\dag \cH_{\r,\r'} \hat{c}_{\r'} .
\end{align}
Here $\hat{c}_{\r}$ removes an atom at site $\r$, 
$\bf{e} \in \{ \pm \hat{\bf{x}}, \pm \hat{\bf{y}}, \pm \hat{\bf{z}} \}$, $\cH$ is the matrix of 
$\hat{H}_{tb}$, and the spin indices have been suppressed on $\hat{c}$, 
$\bm{\sigma}$, and $\cH$; $A$ signifies one of the two 
sublattices of the bipartite division of the cubic lattice.
%into the rock salt structure.  
This model 
%is gapped when $t,t_M \neq 0$, 
and captures the hedgehog character of Zeeman field $\bf{B}_Z$.  
%There is no magnetism on site, but tunneling between sites is modified.  
Starting from sublattice $A$, the barrier to tunneling in the 
$+\hat{\bf{z}}$ direction is higher for an atom with $\sigma^z = +1$ than for 
one with $\sigma^z = -1$, and the barriers to tunneling along $-\hat{\bf{z}}$ 
are interchanged.  These statements also hold for $z\rightarrow x,y$. 
%and $z\rightarrow y$.

The bulk spectrum is given by
\begin{align}
\cH_{\k} &= 2\begin{pmatrix} 0 & g_\k \\ g_\k^\dag & 0 \end{pmatrix}, 
\quad \epsilon(\k) = \pm 2 \sqrt{t^2 f_\k^2 + t_M^2 f_{M\k}} , \notag\\
g_\k &= %e^{i (k_x+k_y+k_z)} 
\!\sum_{j\in \{x,y,z\}}\! (t \cos k_j - i t_M \sigma^j \sin k_j ) ,
\notag\\
f_\k &= \!\sum_{j\in \{x,y,z\}}\! \cos k_j  , \quad
f_{M\k} = \!\sum_{j\in \{x,y,z\}}\! \sin^2 k_j , 
\end{align}
and shown in Figure \ref{fig:tbbulkbands}, where $\cH_{\k}$ is the Fourier transform of $\cH_{\r,\r'}$ and is a matrix in sublattice as well as spin space.  As expected,
it resembles the lowest two (doubly degenerate) bands of the optical
lattice model in Figure \ref{fig:bulkbands}, particularly in that each
band is doubly degenerate at each wave vector.  Note that there is a gap
whenever $t,t_M \neq 0$.  In the deep-well limit, 
the higher bands of Figure \ref{fig:bulkbands} move off to high energies.

The tight-binding model has more symmetry than does %the potential $V$: 
$H_{AF}$:
$\Sigma \cH \Sigma^{-1} = -\cH$, where $\Sigma c_A = c_A$, 
$\Sigma c_B = -c_B$ on sublattices $A$ and $B$.  This is known as ``sublattice'' or ``chiral'' 
symmetry~\cite{Altland1997,*Zirnbauer1996,Ryu2010,*Kitaev2009}, 
which places this model into symmetry class DIII, akin to phase B of $^3$He~\cite{Schnyder2008}.
The associated topological
invariant is particularly straightforward to evaluate~\cite{Ryu2010,Volovik2010,EssinGurarie2011}:
\begin{align}
N_3 &= \pi \int\! \frac{d^3 k}{(2\pi)^3}\, \frac{1}{3!} \epsilon_{abc} 
\,\tr\, \Sigma D^a D^b D^c = 1, \notag\\
D^a &= \cH_{\k}^{-1} \d_{k_a} \cH_{\k} ,
\end{align}
where the integral is over the fcc Brillouin zone. 
%and $\cH_{\k}$ is the Fourier transform of $\cH_{\r,\r'}$, which is a matrix in sublattice as well as spin space:
%\beq \label{eq:Hk}
%\cH_{\k} = 2\begin{pmatrix} 0 & g_\k \\ g_\k^\dag & 0 \end{pmatrix}, \;\;
%g_\k = %e^{i (k_x+k_y+k_z)} 
%\!\sum_{j\in \{x,y,z\}}\! (t \cos k_j - i t_M \sigma^j \sin k_j ) . 
%\eeq

The surface bands of $\hat{H}_{tb}$ are shown in Figures \ref{fig:tbss}
and \ref{fig:tbhs}.  They, too, resemble the corresponding spectra for 
the optical lattice potential.  On the (100) surface the Dirac point
sits in the center of the gap.  More remarkably, on the (111) surface the
disconnected bands that can seen above the upper band in Figure 
\ref{fig:hs} also migrate to the center of the gap in the tight-binding
limit.  That band, once at zero energy, is protected by chiral symmetry
(in this geometry there is only chiral symmetry, so the system is 
formally in class AIII)
and is necessarily flat (the states in that band occupy one of the two sublattices %of the model 
and their
energy is protected by the index theorem as explained, for example, in Ref.~\onlinecite{Gurarie2007},
or by a theorem of Lieb, Ref.~\onlinecite{Lieb1989}). 
Moreover, it can be checked numerically that this flat band
has Chern number 1; this is like the zeroth Landau level of a Dirac mode~\footnote{Note that this disconnected band
violates the usual bulk-boundary correspondence for a TI~\cite{Ryu2010,EssinGurarie2011}.  Our model also has nontrivial 1d chiral invariants for each lattice direction, which are responsible for
this violation, as they require disconnected bands.}.
%It is clear that a flat non degenerate band  
%%%%

\section{``Magnetoelectric'' response in ultracold atomic gases}

There has recently been much discussion of topologically 
nontrivial flat bands as a way to realize fractional quantum Hall physics
without an external magnetic field; typical cases require tuning of
parameters to achieve a very flat band~\cite{Tang2011,*Sun2011,*Hu2011}.  Here the flatness is perfect when
%so long as 
the surfaces respect the sublattice symmetry, with no
%without any 
tuning.

To realize the Hamiltonian Eq.~\rf{eq:topins} or its tight binding version Eq.~\rf{eq:tb} we need to employ atoms with two internal levels,
representing spin,
which can be coupled by a laser. %It seems to us that 
A particularly promising approach would be to use for these two levels
the ground ($^3$S$_0$) and excited ($^3$P$_0$) states of fermionic
alkaline-earth-like atoms such as Sr or Yb; this is attractive due to the 
extremely long lifetime of the excited state and the fact that these states can be coupled directly by an optical laser. 
%operating at an optical frequency.  

The tight-binding Hamiltonian Eq.~\rf{eq:tb} may be created by directly imprinting the tunneling matrices onto the atoms 
 following Ref.~\onlinecite{Jaksch2003,*Dalibard2010}. Alternatively, let us describe realizing the potential of Eq.~\rf{eq:topins}. 
Working with the alkaline-earth-like atoms, the scalar potential and $\sigma_z$
can be realized with lasers at ``magic'' and ``anti-magic'' 
wavelengths~\cite{Dalibard2010}, while $\sigma_x$ and $\sigma_y$ potentials require a laser operating at
the $^1$S$_0$--$^3$P$_0$ transition frequency. Matching the wavelengths of these lasers would require setting up two traveling waves at an angle 
%with each other 
for every standing wave potential. Note that the $^3$P$_0$ state is known to be collisionally unstable. We can eliminate this 
instability if we polarize the nuclear spins of the atoms preventing two $^3$P$_0$ atoms from scattering in the $s$-wave channel. 
$^3$P$_0$-$^1$S$_0$ collisions may also be unstable, although recent experiments indicate that at least in $^{87}$Sr this instability is weak (below experimental sensitivity)~\cite{Ye2011}.

Let us now turn to discussing how to see the 
%nteresting 
physics of this
%, or any other,
%topological insulator 
TI in an optical lattice.  In a crystal, the most 
dramatic consequence is the presence of the surface states displayed 
above.  The topological surface states have a Dirac-like spectrum that 
connects the bulk bands.
%, whereas more generic Shockley states have a 
%parabolic dispersion that does not cross the gap~\cite{Shockley1939}.
While there is some spectroscopic information available for atomic 
gases~\cite{Stewart2008}, a more productive approach may be to look
at macroscopic properties, in particular the response of the gas to 
external forces.

%\section*{Coupling between linear and orbital motion}

Consider atoms of mass $m$ subject to
an additional, constant external force $\bf{F}$. 
%and  which could also rotate with the angular velocity $\bm{\Omega}$.  
%One can generate this 
%rotation via an artificial, orbital magnetic field, which requires 
%coupling to atomic states not considered here.\cite{Dalibard2010}  
%The Hamiltonian is
%%\begin{align}
%%H = \sum_i \left[ H_{AF}(\p_i,\r_i,\s_i) 
%%- \bm{\Omega} \cdot \bf{L}_i - \bf{F} \cdot \r_i \right].
%%\end{align}
%$
%H =  H_{AF} %- \bm{\Omega} \cdot \bf{L} 
%- \bf{F} \cdot {\bf R},
%$
%where %${\bf L}$ is the total angular momentum, 
%${\bf R}$ is the sum of the positions of all the atoms and $H_{AF}$ is the sum
%of Hamiltonians such as Eq.~(\ref{eq:topins}) over all the atoms (plus interaction terms if any). 
The orbital response tensor is 
$
\alpha_j^i = \d \mathcal{L}_j /\d F_i,
%= \frac{\d \mathcal{P}^i}{\d \Omega^j}
%= -\frac{\d^2 \ev{H}}{\d F_i \d \Omega^j},
$
where %$\ev{H}$ is the average energy density and 
$\bm{\mathcal{L}}$ is the average angular momentum density. 
%and $\bm{\mathcal{P}}$ is the average number dipole density. 
%At first glance, 
One expects this quantity to vanish in linear response
when the potential possesses time-reversal and/or inversion symmetry.
However, in a TI with surface $\mathcal{T}$-breaking, it takes the surprisingly large, isotropic value
\beq \label{eq:response}
\alpha_j^i = \pm \frac{m}{h} \delta^i_j, 
\eeq
where $h$ is Planck's constant~\footnote{The solid-state literature that derives this result~\cite{Qi2008,Essin2009,Malashevich2010,*Essin2010} considers charged particles, and in a neutral system one needs to 
divide the constant $e^2/h$ by $e\gamma$,
% identify 
% $\bf{M} = \gamma \bm{\mathcal{L}}$
% %$\bf{P} = e \bm{\mathcal{P}}$, 
% and $\bf{E} = \bf{F}/e$, 
% %and $\bf{B} = \bm{\Omega}/\gamma$, 
with $e$ the charge
and $\gamma = e/2m$ the classical gyromagnetic ratio.}.   This response is very strong. Indeed, applying a force of the order of $E_{\rm r}/d$,
where $d$ is the linear size of the system and $E_{\rm r}= h^2/(ma^2)$ is the lattice recoil energy,  we find from Eq.~\rf{eq:response} that the induced angular momentum is of the order 
$h (d/a)^2$, that is one quantum per %two dimensional 
2D unit cell. That far exceeds what was achieved by rotating the atomic gases directly~\cite{Cornell2004},
recent progress in this effort notwithstanding~\cite{Gemelke2010}.
% In fact, the 
% magnitude $m/h$ of the response is precisely one-half the quantum Hall 
% coefficient, equivalent to $e^2/2h$ in a charged system.  Therefore a 
% quantized orbital polarizability (orbital magnetoelectric 
% polarizability for charge) is a strict three-dimensional analogue of the 
% two-dimensional quantum Hall effect, which is after all a coupling between
% longitudinal and transverse motion.

% It is worth recalling the sign ambiguity.  In fact, the more general
% result is $\alpha_j^i = (2n+1) (m/h) \delta^i_j$ for any integer $n$.  In
% other words, the bulk physics only determines the fractional response in
% terms of the quantum $2m/h$; the precise value is set by details of the 
% surface.  Typically, one assumes that the system adopts one of the minimal
% values $\pm m/h$, since any extra part can be interpreted as an extra 
% ``quantum Hall layer'' residing at the surface.

The striking signature of the TI phase in a gas of 
fermions should then be 
%a polarization of the cloud in the presence of an
%externally imposed rotation, or 
a rotation of the cloud in response to
a linear potential gradient, if there were some $\mathcal{T}$-breaking present.  In the present case, 
%Moreover, any realization of our model would include 
a parabolic trap will necessarily break all the relevant symmetries
(both chiral symmetry if present and $S$, since it involves translations),
enabling a strong response. 
%As a result, 

\begin{figure}
\begin{minipage}[c]{0.45\columnwidth} \subfigure[]{ \includegraphics[width=\columnwidth]{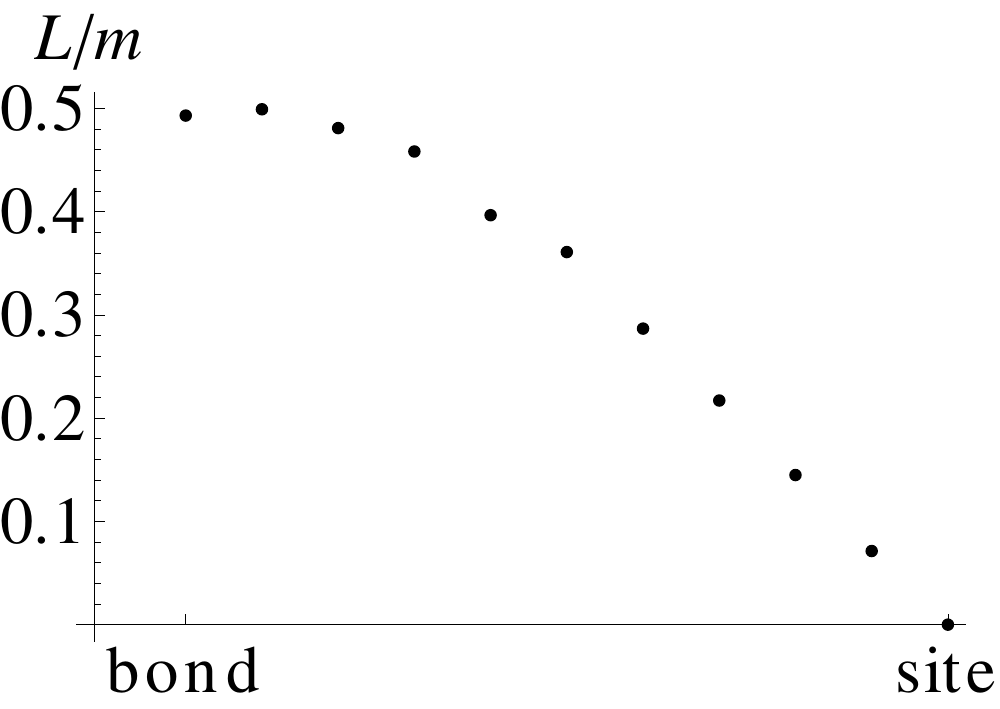}
\label{fig:netcirc}
} \end{minipage} 
\begin{minipage}[c]{0.45\columnwidth} 
\subfigure[]{ \includegraphics[width=\columnwidth]{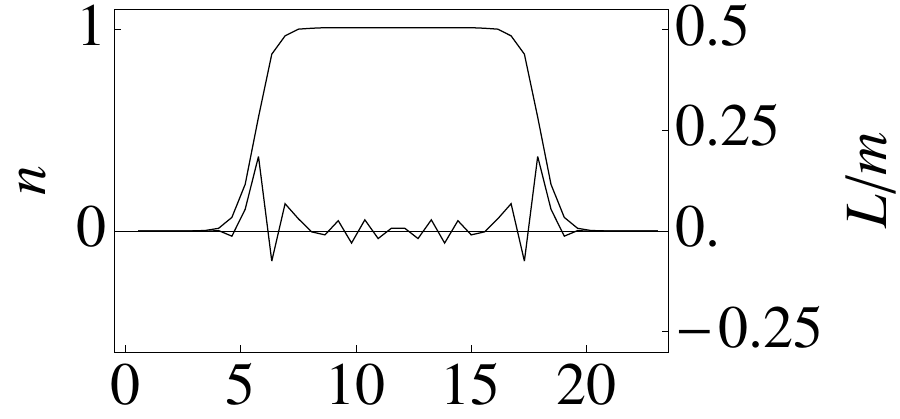}
\label{fig:bondplot}
} \\ \vspace*{-0.1in}
\subfigure[]{ \includegraphics[width=\columnwidth]{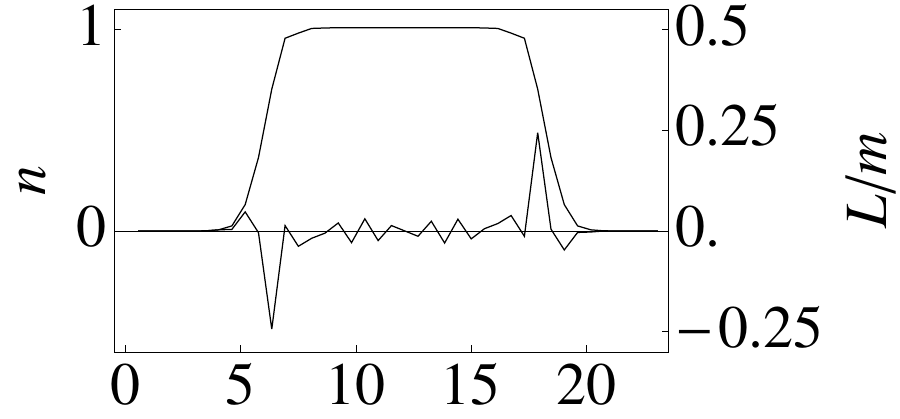}
\label{fig:siteplot}
} 
\end{minipage} 
\caption{ \label{fig:circ}
(a) Net circulation $L/m$ per two-dimensional unit cell in a harmonic 
trap, in units of $ta^2/h$, for $t_M = t$, $m \omega^2 a^2 = t/4$, 
$\mu = 1.5 t$.  The 
abscissa gives the position of the trap center relative to the lattice.  Also shown are the circulation $L/m$ (lower curve) 
and particle number per site $n$ (upper) as a function of position (in units of $a$) for (b) a bond-aligned and (c) a site-aligned trap.
}
\end{figure}

In fact, {\it the mere presence of the trap induces rotation} %of the system 
%(except for highly symmetric positions of the trap center where the rotation may vanish). 
in general; after all, shifting the trap by $\Delta s$ is essentially 
equivalent to a force $m\omega^2 \Delta s$, for trap frequency
$\omega$.  We have computed the circulation for a gas tightly confined 
in the [111] direction by a harmonic trap; see Fig.~\ref{fig:circ}.  
To make 
this numerically tractable we have imposed periodic boundary conditions (we do not expect a weak, parabolic 2D trap to change any resulting physics), and
computed the circulation per 2D unit cell for $\hat{H}_{tb}$~\cite{Ceresoli2006,*Thonhauser2005,*Shi2007,*Xiao2005,*Lopez2011,*Thonhauser2011},
\begin{equation}
\frac{L}{mN_sa^2} = \frac{\pi}{ih} \int\! \frac{d^2k}{(2\pi)^2}\, \epsilon_{ab}\,
\tr\, (P-Q) (\d^a P) \tilde{\cH} (\d^b P),
\end{equation}
on a 100-by-100 grid of the (111) Brillouin zone.  Here $\tilde{\cH}$ is the two-dimensional Fourier 
transform of $\cH_{\r,\r'} + m\omega^2 s^2/2 - F s - \mu$, with $s$ the [111] coordinate; $P$ ($Q$) projects onto the negative- (positive-) energy states of $\tilde{\cH}$; and $N_s$ is the number of surface unit cells.
The sign of $L$ changes for alternate bonds, since the $\mathcal{T}$-breaking term switches sign.
The derivative $|\d (L/Na^3)/\d F| \lesssim 0.5 m/h$, in agreement with Eq.~\rf{eq:response}, when the trap is aligned with
lattice sites; here $N$ is the particle number, and provides a suitable measure of the size of the trapped system since the bulk density is one particle per site.  
% The magnitude of the rotation is no longer quantized, but its maximum roughly corresponds to the response given by Eq.~\rf{eq:response}. 
% Fig.~\ref{fig:circ} shows the numerically calculated response of the model Eq.~\rf{eq:tb}  that is tightly confined in the [111] direction by a parabolic potential (for reasons of simplifying the calculation we assumed periodic boundary conditions in the perpendicular directions, although we do not expect a weak, parabolic 2D trap to change any resulting physics) following  Ref.~\onlinecite{Ceresoli2006,*Thonhauser2005,*Shi2007,*Xiao2005}. 
The rotation takes its maximum when the trap is aligned with the bond centers, 
%and the maximum magnitude of the angular momentum conforms to the estimate after Eq.~\rf{eq:response} as shown in Fig.~\ref{fig:netcirc}. 
and the circulation is concentrated mainly at the surfaces,
%edges of the insulator, 
as is clear from Figs.~\ref{fig:bondplot} and \ref{fig:siteplot}.

The resulting rotation may be measured with a variety of methods: 
for example,
by switching off the lattice and observing a flattening of the rotating
cloud; 
by measuring the way the Fourier-transformed density distribution 
$n_\k$ vanishes at $\k = 0$; 
by measuring the frequency modes of the rotating 
gas~\cite{Dalibard2000,*Cornell2001};
or by Bragg spectroscopy~\cite{Ketterle1999}.

Note that the response tensor $\alpha$ is well-defined even in the 
presence of interactions.
%, unlike the topological invariant we computed
%based on inversion eigenvalues.  
The invariant for the chiral limit is 
similarly well-defined with interactions, using
%when evaluated in terms of the 
Green's function rather than the single-particle Hamiltonian~\cite{Gurarie2011}.
That is, the TI phase should be stable to the 
introduction of weak interactions.  It would be interesting to check this
result directly in cold Fermi gases, which display Feshbach resonances 
that lead to tunable interactions (although that might require using alkali atoms coupled by Raman transitions instead
of alkaline earths as outlined above); it would be even more interesting to
study the Mott-type phases that should emerge upon introducing very 
strong interactions~\cite{Pesin2010}.
% that localize the particles but not the spin 
% degree of freedom~\cite{Pesin2010}.

Finally, it is worth pointing out connections to other systems that 
display interesting behavior in the presence of magnetic textures.  In 
particular, such textures have been argued to be give an important 
contribution to the anomalous Hall effect in ``colossal 
magnetoresistance'' materials~\cite{Ye1999} and in MnSi~\cite{Binz2006}. 
Such textures can produced with a Zeeman field such as of Eq.~\eqref{eq:topins}, when it is incommensurate with the lattice or when the lattice
is absent entirely.

\acknowledgements

We thank E.~A.~Cornell, C.~Vale, P.~Drummond, M.~D.~Swallows, and especially A.~M.~Rey for discussions concerning the realization and observation of topological insulators.  AME thanks R.~S.~K.~Mong and J.~E.~Moore for earlier collaboration on AFTIs.  AME is supported by DOE award de-sc0003910 and VG by NSF grants PHY-0904017 and PHY-0551164. VG is also grateful to KITP Santa Barbara and the physics department of the University of Melbourne where part of this work was done.

\bibliography{AFTI}

\end{document}